# Comparison of Geant4-DNA and RITRACKS/RITCARD: microdosimetry, nanodosimetry and DNA break predictions

Victor LEVRAGUE[1], Rachel DELORME[1], Mathieu ROCCIA[1], Ngoc Hoang TRAN[2], Sebastien INCERTI[2], Michael BEUVE[3], Etienne TESTA[3], Ianik PLANTE[4], Shirin RAHMANIAN[5] and Floriane POIGNANT[5]

[1]Univ. Grenoble Alpes, CNRS, Grenoble INP, LPSC-IN2P3, 38000 Grenoble, France
[2]Université de Bordeaux, CNRS/IN2P3, LP2i, UMR 5797, 33170 Gradignan, France
[3]IP2I, Université Claude Bernard Lyon 1, CNRS/IN2P3, 4 rue Enrico Fermi, 69622 Villeurbanne Cedex, France
[4]KBR, Houston, TX 77058, USA
[5]Analytical Mechanics Associates Inc., Hampton, VA 23666, USA

**Abstract** This work aims at investigating the impact of DNA geometry, compaction and calculation chain on DNA break and chromosome aberration predictions for high charge and energy (HZE) ions, using the Monte Carlo codes Geant4-DNA, RITRACKS and RITCARD. To ensure consistency of ion transport of both codes, we first compared microdosimetry and nanodosimetry spectra for different ions of interest in hadrontherapy and space research. The Rudd model was used for the transport of ions in both models. Developments were made in Geant4 (v11.2) to include periodic boundary conditions (PBC) to account for electron equilibrium in small targets. Excellent agreements were found for both microdosimetric and nanodosimetric spectra for all ion types, with and without PBC. Some discrepancies remain for low-energy deposition events, likely due to differences in electron interaction models. The latest results obtained using the newly available Geant4 example "dsbandrepair" will be presented and compared to DNA break predictions obtained with RITCARD.

## 1 Introduction

At the cellular scale, energy deposition events induced by low linear energy transfer (LET) ionizing radiation create biological damage including DNA double strand breaks (DSB) that are relatively homogeneously distributed. On the contrary, high LET radiation, such as high charge and energy (HZE) ions, are well known to induce heterogeneously distributed energy deposition events that can create complex biological damage, including clustered DSB. Clustered DSB are critical damage events difficult to repair, leading to increased cell death and chromosome aberrations (CAs) compared to low LET radiation exposure. This increased biological effectiveness makes high LET ions of great interest in medicine for increasing the therapeutic efficacy of cancer treatment, an advantage utilized in hadrontherapy. However, exposure to HZE ions also comes with carcinogenesis risks, which is an issue for both hadrontherapy (secondary cancer) and long-term space travel. Cellular biological mechanisms driving cancer initiation, promotion and progression for high-LET exposure remain to be fully elucidated[1] and constitute an active field of research. In that context, Monte Carlo track structure simulations are valuable tools to identify key cellular events and quantify the mechanisms.

The track structure code RITRACKS (Relativistic Ion Tracks) [Plante 2008], together with the code RITCARD (Relativistic Ion Tracks, Chromosome Aberrations, Repair, and Damage) [Plante 2019a], allow to simulate the transport of HZE ion tracks in micro and nanostructures. The yield of DNA damage is determined by calculating energy deposited in nanovoxels containing DNA, without details of the DNA geometry. Estimation of CA yields are obtained by modeling DSB (mis)repair and classifying the aberrations. However, DNA geometry influence the yield of DNA breaks. Previous numerical studies have shown that chromatin compaction decreases the yield of DNA damage generated by ionizing radiation, with dependence on ion charge and energy [Tang 2019]. Such changes in DNA damage yield are likely to impact CA yields and are thus important to consider in DNA damage and repair simulations.

We intend to use highly detailed DNA geometries available in the Geant4-DNA toolkit to compute DNA breaks for different level of DNA compaction and combine the results with RITCARD to determine the impact of DNA compaction on CA predictions. A first necessary step to fulfill this objective is to ensure the consistency of radiation transport models used in RITRACKS and Geant4-DNA.

The work that will be presented corresponds to this first phase of the project. We compared energy deposition at micrometric and nanometric scales for different ions with LET ranging from ~0.4 keV/μm up to 235 keV/μm. The calculation of DNA breaks and their complexity with the new calculation chain available in Geant4 11.2 is an ongoing work, and comparison with RITCARD will be presented based on availability at the time of the conference.

## 2 Materials and Methods

### 2.1 Geant4-DNA

Geant4-DNA is an open-source code developed through an international collaboration, that was originally initiated in the context of space radiation and extended to medical physics applications. It includes physics processes that allow event-by-event particle transport down to very low energy (typically down to 7 eV for electrons). It includes different geometric models of DNA for a full human cell nucleus with a detailed atomic-scale structure of the molecule[2] [Incerti 2018, Kyriakou 2022]. This precision allows to account for damage coming from both direct ionization of the DNA by ion tracks and secondary electrons, and the attack of chemical species on DNA, produced by the radiolysis of the surrounding water.

---
[1] https://humanresearchroadmap.nasa.gov/

The presented results were obtained with the latest version of Geant4 (v11.2) using DNA-physics option 2, for which the model of heavy ion transport (*G4DNARuddIonisationExtendedModel)* was significantly updated. Calculations of the energy distribution of secondary electrons induced by water-ion ionization showed that Geant4 v11.2 results in a much better agreement with RITRACKS compared to previous Geant4 versions. In addition, developments were made in our Geant4 example to include periodic boundary conditions (PBC), based on previous work[3], as a solution to account for energy deposition due to delta electrons generated by distant tracks, simulating electronic equilibrium in small targets. The PBC is applied to electrons. We have also implemented an efficient nano-voxel mesh scorer compatible with the PBC option in order to compute the nanodosimetry spectra. Finally, we modified the physics list to extend the maximum energy limit for the *Rudd Ionisation Extended Model* from 300 MeV/n to 1 GeV/n for ions. We also modified the physics list to use the *Rudd Ionisation Extended Model* for proton and alpha particles. For alpha particles, cross sections were calculated to extend them from 100 MeV/n up to 1 GeV/n. This allows a consistent use of the Rudd ionization model for all ions.

The Geant4-DNA version 11.2 also contains two new DNA break calculation chains, called "*moleculardna*" and "*dsbandrepair*" that allows to calculate early radiation-induced DNA damage, such as DSBs and to model repair processes. The *dsbandrepair* application also contains different atomic scale models of DNA, including both the radioresistant heterochromatin (HC) compact form and the radiosensitive euchromatin (EC) less compact form. In a second step of this benchmark process, we intend to use this example to compare results of simple and complex strand break yields in our irradiation configurations, using both EC and HC DNA forms and compare it with the RITRACK/RITCARD DSB predictions. This is an ongoing work that will not be described further in this abstract.

## 2.2 RITRACKS/RITCARD

RITRACKS/RITCARD has been developed by NASA, focusing on space radiation applications. RITRACKS is a Monte Carlo track-structure code that simulates the full transport of protons and high charge and energy particles using the Rudd model, as well as the secondary electrons down to very low energy, as detailed elsewhere [Plante 2008, Plante 2011a]. Recent developments include the possibility to activate PBC in micrometric volumes to account for energy deposition in targets by energetic delta electrons generated by distant tracks. RITRACKS can calculate energy deposition in nanovoxels as well as in spherical and cylindrical micrometric targets [Plante 2011b, Plante 2021]. Contrary to Geant4-DNA, RITCARD calculates DNA DSBs based on energy deposited within nanovoxels containing 2,000 kbp of DNA, encompassing both the direct and indirect effects without going down to details at the atomic scale. RITCARD also includes repair processes which can account for dose-rate effect. Finally, RITCARD allows to classify the damage events into CA, an option not yet available in Geant4-DNA.

## 2.3 Micro- and nanodosimetry spectra

We calculated single track energy deposition distributions in micrometric and nanometric targets for several monoenergetic ion beams. The energy imparted to the target, $\varepsilon$ (in eV or J) is,

$$\varepsilon = \sum_i \varepsilon_i,$$

where the summation over $\varepsilon_i$ represents all the energy deposition events (ionizations/excitations). $f_1(\varepsilon)$ is defined as the probability distribution of $\varepsilon$ for a single ion track generating energy deposition events in the target. Comparisons were made for the following ions: $^1H^+$ (150 and 250 MeV), $^4He^{2+}$ (250 MeV/n), $^{12}C^{6+}$ (290 MeV/n), $^{16}O^{8+}$ (55 and 350 MeV/n), $^{28}Si^{14+}$ (170 MeV/n), $^{48}Ti^{22+}$ (300 MeV/n) and $^{56}Fe^{26+}$ (300, 450 and 600 MeV/n). These ions were chosen because of their interest in hadrontherapy and space research, and experimental chromosome aberration data are available [Hada 2014, Plante 2019a, Plante 2019b]. The simulation set up for microdosimetry calculation is described in Figure 1. For both microdosimetry and nanodosimetry calculations, the geometry was set as a water box of 12 µm in the longitudinal direction (direction of ion transport), and 16 µm x 16 µm in transverse directions. It is worth mentioning that with the PBC option activated, when a delta-electron is re-injected in the world volume, it is considered as a new track, as it mimics contribution from distant tracks. To mimic particle equilibrium, PBC were applied as followed: when an electron passed through one side of the irradiated volume, it appeared on the opposite side with the same velocity vector. The irradiated box contained a micrometric target which was a centered ellipsoid volume with a transverse radius of 7.1 µm and a longitudinal radius of 2.5 µm, representing the cell nucleus of a human fibroblast in an adherence configuration[4]. The nanometric targets were cubical voxels of 75 nm side that were distributed within the cell nucleus, which aims at representing a fragment of DNA. Single ion energy distribution $f_1(\varepsilon)$ were calculated using a logarithmic binning (1,000 bins) between 6 eV to $2 \times 10^6$ eV for microdosimetry and 6 eV to $2 \times 10^4$ eV for nanodosimetry. For microdosimetry, results are normalized so that the first moment of $f_1(\varepsilon)$ gives the average energy deposited by one track to a target. For nanodosimetry, results were normalized so that $\int_0^\infty f_1(\varepsilon)\, d\varepsilon = 1$.

---

[2] http://geant4-dna.org/
[3] https://github.com/amentumspace/g4pbc
[4] https://geant4-dna.github.io/molecular-docs/

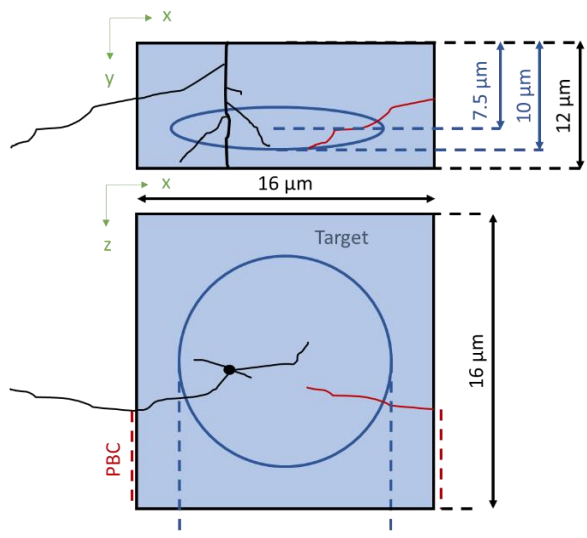

*Figure 1. Simulation set-up for microdosimetry calculation. The light blue volume represents the simulated water volume, the dark blue circle the microdosimetry target, the black lines an ion track, the red line an electron that exited the volume and for which PBC were applied.*

## 3 Results

Figure 2 shows the microdosimetry results obtained with Geant4 11.2 and RITRACKS with activated PBC for 9 different ions. Figure 3 shows the nanodosimetry results obtained with Geant4 11.2 and RITRACKS with activated PBC for 9 different ions. Excellent agreements were found for both microdosimetric and nanodosimetric spectra for all ion types and energies, with and without PBC *(without PBC are not shown)*. Microdosimetry spectra differ mainly for low-energy deposition events below 100 eV, especially for He, C and O ions, likely due to differences in electron interaction models. In the nanodosimetry spectra, slight discrepancies are observed in the 100 eV-10 keV $\varepsilon$ range for heavy ions. These results will be presented at the conference with a deeper analysis on the comparisons.

## 4 Discussion and Conclusion

The present work aims at investigating the impact of DNA geometry on DNA break and CA predictions for HZE ions using the codes Geant4-DNA, RITRACKS and RITCARD.

The first phase of the comparison process showed excellent agreements in microdosimetry and nanodosimetry spectra. This allows us to validate the Geant4-DNA newly released *Rudd Extended Ionisation Model* for heavy ion transport by comparing it with RITRACKS, that has been previously validated on several experimental data for HZE ions. In addition, we included changes in the Geant4 *Rudd Extended model* to also allow its use for alpha particles beyond 100 MeV/n. Finally, practical improvements were made to allow the use of PBC and registering physical data in both microstructures and nanovoxels, improvements that future Geant4 users may benefit from. The second step of code comparison between Geant4 and RITRACK/RITCARD is ongoing and will be dedicated to DNA break comparisons. It will focus on the impact of DNA geometry, compaction, and detailed direct/indirect effect consideration for DSB, and how such changes impact CA yields.

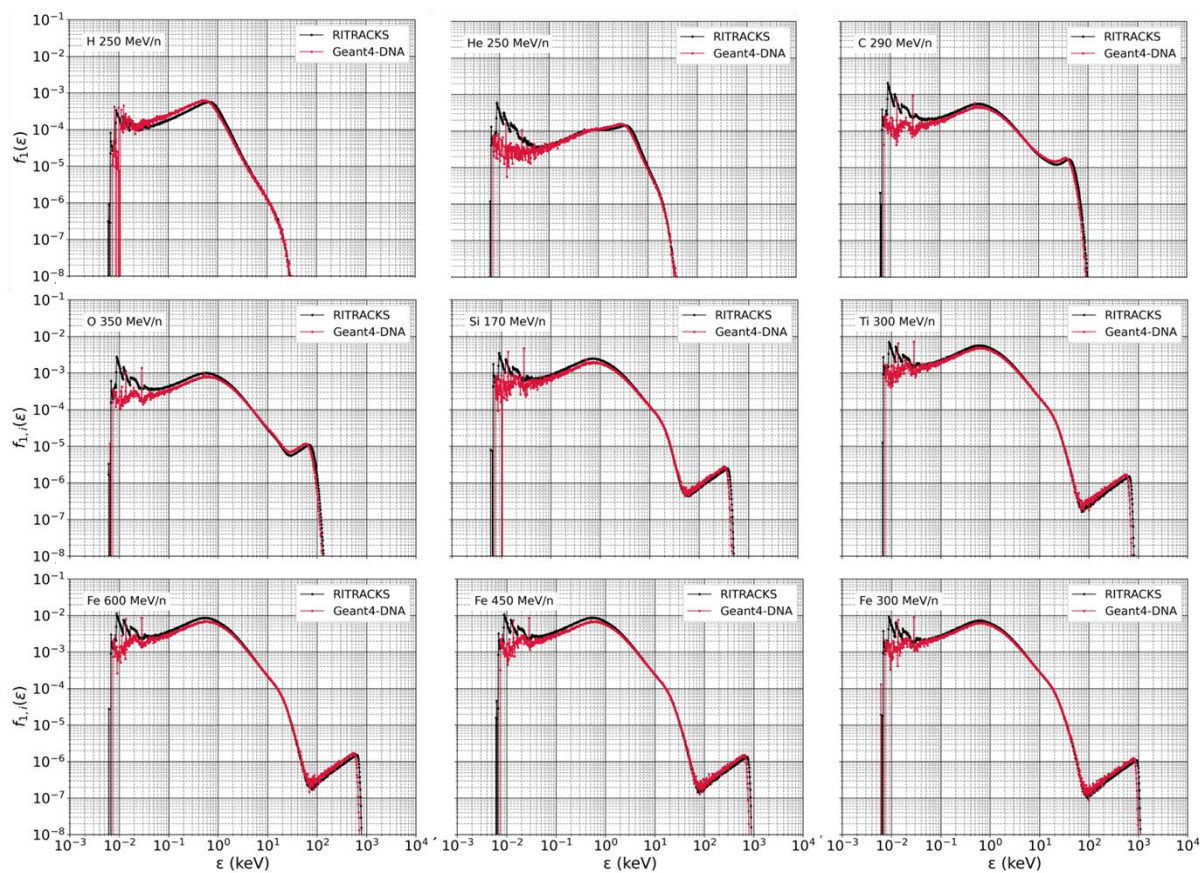

*Figure 2. Comparison of single ion energy distribution in micrometric targets between RITRACKS (black) and Geant4-DNA (red).*

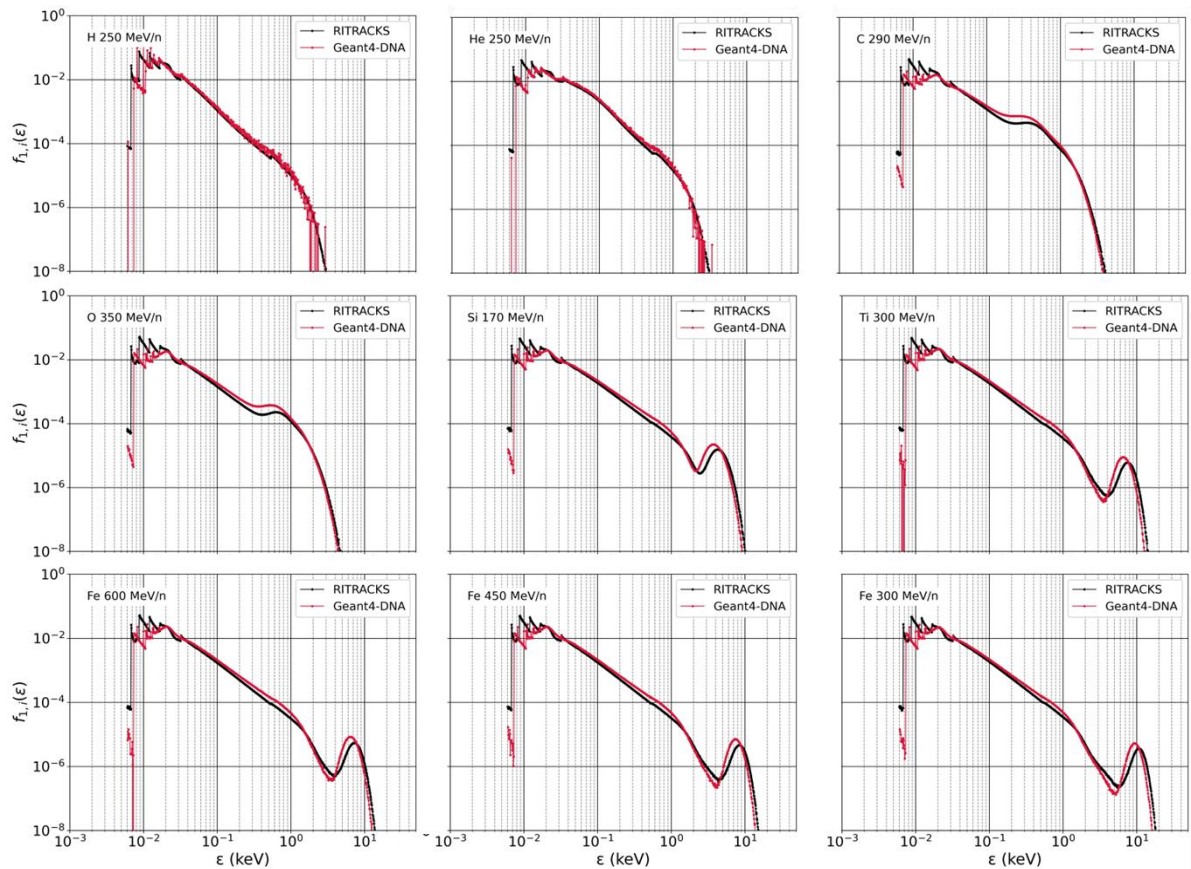

*Figure 3. Comparison of single ion energy distribution in nanometric targets between RITRACKS (black) and Geant4-DNA (red).*